# Vortex stability and permanent flow in non-equilibrium polariton condensates

G. Tosi[1], D. Sanvitto[1], M. Baudisch[1], E. Karimi[2], B. Piccirillo[2], L. Marrucci[2], A. Lemaître[3], J. Bloch[3], and L. Viña[1]

[1]Departamento de Física de Materiales, Universidad Autónoma de Madrid, Madrid 28049, Spain,

[2]Dipartimento di Scienze Fisiche, Università di Napoli Federico II and CNR-SPIN, Napoli 80126, Italy,

[3]LPN/CNRS, Route de Nozay, 91460, Marcoussis, France.

**Abstract.** We study the effects of imprinting a single-quantized vortex on the steady state of a microcavity exciton-polariton condensate generated via parametric scattering. Interestingly we observe two distinct regimes: in the first case, at low polariton densities, the effect of the pulsed probe, containing the vortex state, is to generate a gain response in the condensate lasting for tens of picoseconds during which no dissipation of the circulating currents is detected. In the second regime, at higher densities, the gain lasts much less and the circulation is imprinted directly into the steady state, which acquires permanent rotation for as long as the vortex remains within the condensate. We use two different ways of measuring the circulation of the condensate and demonstrate that in both cases, polariton condensation in the parametric scattering regime can sustain permanent supercurrents.

**I. INTRODUCTION**

Microcavity polaritons are quasiparticles constituted by the strong coupling between a cavity-photon and an excitation in a semiconductor material: usually excitons in quantum wells [1]. Since both, excitons and photons, are bosonic particles, also polaritons obey Bose-statistics, and, under sufficiently high density, collective coherent phenomena associated to the Bose-Einstein condensation arise [2, 3]. One of the very peculiar features of polariton particles is their extremely light mass, which makes them perfect candidates for high critical temperature condensates, as recent experiments suggest [4]. However due to their dissipative nature (some of the longest lifetimes of polaritons in a cavity are around 15 $ps$ [5]) they belong to a new class of condensates which undergo a phase transition despite their non-equilibrium character, which makes them an extremely interesting testbed for the observation of new phenomena associated to this kind of out of equilibrium bosonic particles. Different manifestations of superfluidity, such as diffusion-less flow without resistance [6], density and momentum dependent signatures of sub- and supersonic flow [7], stability for higher order of circulation [8] and the presence of different imprinting regimes of vortex states, reported in this manuscript, being some of the examples of the plethora of new phenomena of non-equilibrium polariton condensates.

In this paper we show that a state of condensed polaritons generated under optical parametric scattering (OPO) regime can be perturbed by a pulsed laser probe carrying a single-quantized circulation of angular momentum which, depending on the coherence of the polariton steady state, can: either only amplify the emission – in this case the vortex state lasts as long as the gain is sustained in the system –, or imprint the circulation onto the polariton condensate as another metastable flow pattern sustained by the same boundary conditions as those of the vortex-less state [9]. This latter observation provides the best way of defining superfluidity for driven dissipative systems [10]. To demonstrate circulation we study the peculiar phase pattern obtained by interfering the emission of the polariton state with a reference wave of constant phase. The visibility of the vortex singularity shows that the circulation is not lost in both regimes. Moreover, in the case of an imprinted vortex, we demonstrate that phase winding is also associated with a difference in wavevector between polaritons circulating on opposite sides of the vortex.

**II. EXPERIMENT**

The sample studied is a $\lambda/2$ AlAs microcavity with a $20\ nm$ GaAs quantum well placed at the anti-node of the cavity electromagnetic field. All experiments are done in an optical cryostat, maintained at $10\ K$. The normal mode splitting (Rabi splitting) was $4.4\ meV$ and a slightly negative detuning (between 1 and 3 meV) was chosen for the experiments.

The sample is resonantly pumped around the inflection point of the lower polariton branch (LPB) by a continuous wave (CW) Ti:sapphire laser with momentum $k_p$ and energy $E_p$ as shown in Fig. 1(a). The power of this laser is always set above threshold ($250\ W \cdot cm^{-2}$) for the generation of a condensed polariton state around $k_s = 0$ and $E_s$ (called signal) and another weaker state (idler) at $k_i = 2k_p - k_s$ and energy $E_i = 2E_p - E_s$.

After the parametric scattering process is set the signal polariton emission is collected and sent to a spectrometer-streak camera system for time and space resolved images of the whole area covered by the pump ($\sim 2x10^4\ \mu m^2$). An image of the signal emission is shown in Figure 1(b) for a pumping power 6 times the threshold. At a given time the signal is perturbed by a pulsed Laguerre-Gauss laser beam, created by scattering a pulsed Ti:sapphire Gaussian beam in a hologram with a fork-like dislocation in its diffraction pattern, lasting for $2\ ps$ and in resonance with the signal polariton condensate [Fig. 1(b-d)]. Real space images can be also mixed in a Michelson interferometer with a reference wave (in our case an expanded region of the signal) to obtain phase dependent intensity-patterns which unravel the $2\pi$ dislocation around the vortex core. This point is characterized by the typical fork-like changes in the number of fringes, as shown in Figure 1(c).

Due to the very low intensity of the emission and the very fast detection times, time-resolved images are averaged over billions of shots. Therefore we are only able to measure deterministic (not random) dynamics of the system.

In order to highlight the spatial density modifications and its evolution introduced by the probe, we remove (unless otherwise stated) the background steady state emission in such a way that negative values of intensity mean a decrease in density introduced into the steady state.

**III. RESULTS**

The behavior of the vortex injected by the probe laser depends strongly on the excitation condition of the driving field (pump laser). At low excitation powers (close to threshold values), as presented in Figure 2 (a-c), the polariton signal is strongly perturbed by the probe laser which activate a long-living parametrically amplified state on top of the steady state of the unperturbed signal. This regime is characterized by a long living gain, fed by the pump field, which lasts for times much longer than the bare polariton lifetimes. A streak camera image, shown in Figure 3(a) display the vortex state persisting for at least $80\ ps$ after the probe has arrived. During this time we have observed that no dissipation takes place on the circulating current despite the decay of the gain population, which is demonstrated by a high visibility of the fork dislocation in the interference images [see Fig. 4 (a)] down to the last detectable gain signal [11]. However, no effect is observed in the signal steady state after the extra polariton population has disappeared (Fig. 3(a)).

On the other hand, under high CW pump powers the probe gain is very short, lasting in the microcavity only slightly more than the pulse. However, under these conditions, the vortex carried by the probe laser field is imprinted into the polariton steady state of the signal and is observed to persist for more than a hundred picoseconds − normally only limited by the time at which the vortex leaves the spatial extension of the condensed signal. A typical streak camera image, obtained under this regime, is shown in Figure 3(b). Here it is clearly visible the appearance of a short gain, immediately lost after a few tens of picoseconds, coexisting with a depression in the region of the vortex core [see also Figures 2(d-f)]. This dip is usually $\sim 10 - 20\%$ of the signal intensity and moves around before, eventually, being expelled out of the condensate. By interfering this images with an expanded part of itself (which works as a constant phase reference beam), we can follow the corresponding time-evolution of the interference fringes. In this manner, we observe that the deep core introduced by the probe vortex is always associated with a fork-like dislocation, and we can conclude that the signal state sustains the metastable permanent flow introduced by the pulse. Note that while the power of the driving field is crucial for the modification of the steady state, the probe power has little effect once a minimum threshold is crossed.

Figure 5 shows two snapshots of the imprinted vortex $132\,ps$ after the probe arrival. The real space image of the signal emission show the vortex core, in the deepest part of the image, surrounded by regions of varying emission intensity, mainly due to a modulated background given by imperfections and fluctuating photonic or excitonic potential. As a consequence the vortex is constrained by some potential profiles and forced to follow always a similar path along the way out from the polariton signal state. This is one of the main reason for the strong visibility of the fork [Fig. 5 (b)], although every image is an average over different experiments starting with the same initial conditions. Despite this deterministic behavior, it is certainly possible that some percentage of the experiments follow a random walk leading to a reduction of the visibility of the imprinted vortex.

The physical mechanism underlying the transition between the two regimes can be explained in terms of the spatial coherence of the steady state. In Figure 6 we show the interference images of the OPO signal with its own expanded part for two different pump powers, one close to the condensation threshold (a) and the other one 6 times bigger than it (b). The visibility of the interference fringes is a measurement of the long-range order coherence of the signal. Under low excitation power, the OPO signal has low long-range order coherence (Fig. 6(a)). This fact can lead to a loss of the phase winding imprinted by the probe and the signal cannot sustain any permanent flow characteristic of superfluids. On the other hand, increasing the pump power, the signal exhibiting a more coherent state (Fig. 6(b)) can display superfluid phenomena: the imprinted vortex survives as a metastable permanent flow in the OPO signal steady state. In this sense, as already pointed out in ref. 10, the superfluid behavior is characterized by a dramatic increase in the lifetime of a quantized vortex.

Note that an external probe cannot equally easily transfer its angular momentum in every point of the steady state signal. This is possibly due to the supercurrent pattern generated by the OPO itself and by the potential landscape caused by defects, dislocations and cavity mirror fluctuations, which favor certain region rather than others.

Another way of measuring the rotation of the polariton supercurrents is probing the difference in momentum at the two opposite sides of the vortex core, as proposed in ref. 10 and depicted in Figure 4(b). If the polariton fluid is circulating, the momentum of these two sides should have opposite directions, and would result in a finite difference in their momenta ($\Delta k$), which corresponds to the average angular momentum with respect to the vortex core. In Figure 4(c) the momentum distribution corresponding to the regions marked with ↑ (↓) are depicted by a red (black) line, respectively. We have observed a finite difference in the momentum distribution between the left and right far field spectra of the signal polaritons for times much longer than the extra population lifetime [Fig. 4(d)]. After $50\,ps$ the difference falls to zero due to the movement of the vortex core outside the central region. This again demonstrates that the signal steady state holds a permanent rotation long after the amplified gain population is gone.

## IV. CONCLUSIONS

We have demonstrated that it is possible to give origin to a permanent rotation of polariton condensates by injecting a vortex state by an external pulsed laser beam into a steady state of polaritons condensate created via optical parametric scattering. Depending on the pumping conditions and the position on the sample, the vortex is created just in the triggered population or, in the other case, transferred as a metastable new state to the steady state of the signal. The vortex sustained by the steady state is the evidence of the manifestation of the superfluid behavior in non-equilibrium polariton condensates.

**ACKNOWLEDGEMENTS**

We are indebted to F. M. Marchetti, M. Szymanska, C. Tejedor and F. P. Lassy for helpful discussion and careful reading of this manuscript.

This work was partially supported by the Spanish MEC (MAT2008-01555 and QOIT-CSD2006-00019), the CAM (S2009/ESP-1503) and FP7 ITNs `Clermont4' (235114). D.S. acknowledges financial support from the Ramón y Cajal program. G.T. is grateful for the FPI scholarship from the Ministerio de Ciencia e Innovación.

**FIGURE CAPTIONS**

**FIGURE 1.** Experimental scheme. (a) OPO dispersion with the idler, pump and signal states placed at the LPB. (b) OPO spatial image with the pulsed vortex at its time of arrival. (c) Interference fringes with a fork-like dislocation and its (d) corresponding spatial image of the pulsed vortex.

**FIGURE 2.** Vortex imprinting regimes. (a) and (d) Steady state for low and high pump powers, respectively, before the probe arrival time. For each of both cases, it is possible to follow the time evolution of, (b) and (e), the extra rotating population created after the arrival of the pulse and, (c) and (f), its subsequent evolution. The four arrows indicate the position of the vortex core. Note that in these images the steady state emission has not been subtracted. The colors in the figures and their correspondent values at the colorbars show that, at low pump power, the density of the steady state is small or even close to zero, and, at high pump power, the steady state has a higher density in the whole region and a spatial structure. Also, by looking at the images after the arrival of the pulse, we note that in the first case there is no decrease of density in the region of the vortex core, which means that the vortex is not imprinted; in the second case there is a decrease of density in the region of the core even when the extra population has disappeared, which means that the vortex is imprinted in the steady state.

**FIGURE 3.** Vortex imprinting regimes (cross sections of the vortex time evolution – in both images the steady state is subtracted). (a) The vortex dies with the extra population (there is no negative values in the region on the core). (b) The vortex is imprinted into the steady state (the negative values of intensity show that there is a core in the steady state even when the extra population dies).

**FIGURE 4.** Two different ways of measuring the rotation of the supercurrents. (a) Interference fringes with a fork dislocation characteristic of a single-quantized vortex. To calculate the amount of rotation we take the intensity profile of a section that crosses the extra fringe and the minimum between the two fringes on the opposite side. The visibility between these two points gives the amount of circulation. (b) Another way of measuring the rotation: the momenta of two opposite sides from the vortex core have contrary directions due to the circulation. (c) Wavenumber distributions - red and black lines - corresponding to each of the two regions marked on figure (b) - right and left, respectively -, showing a small and opposite deviation from zero, which characterizes the circulation. (d) Time evolution of the difference in wavenumber ($\Delta k$) of the two opposite sides (red dots) as opposed to the extra population introduced by the probe (black

line). The difference in wavenumber $\Delta k$ is taken by subtracting the peak value of the wavenumber distribution of one side of the vortex from the peak value corresponding to the opposite side. Note that the circulation persists longer than the extra population, until the time, given by the dashed blue line, when the vortex leaves out the region of measurement.

**FIGURE 5.** Metastable vortex imprinted into the steady state (the steady state is subtracted). (a) Penetration of the vortex into the steady state evidenced by the negative values of its core. (b) Interference image corresponding to (a) showing the fork-like dislocation at the region of the core (dashed lines are guides-to-the-eye).

**FIGURE 6.** Coherence of the OPO signal state as a function of the pump excitation power. At low powers (close to the condensation threshold) the signal presents low long-range order coherence, and low visibility fringes are seen. At high power the signal shows more coherence and fringes can be clearly observed. The images are obtained by interfering the OPO-only signal with an expanded part of it in a Michelson interferometer.

**FIGURE 1**

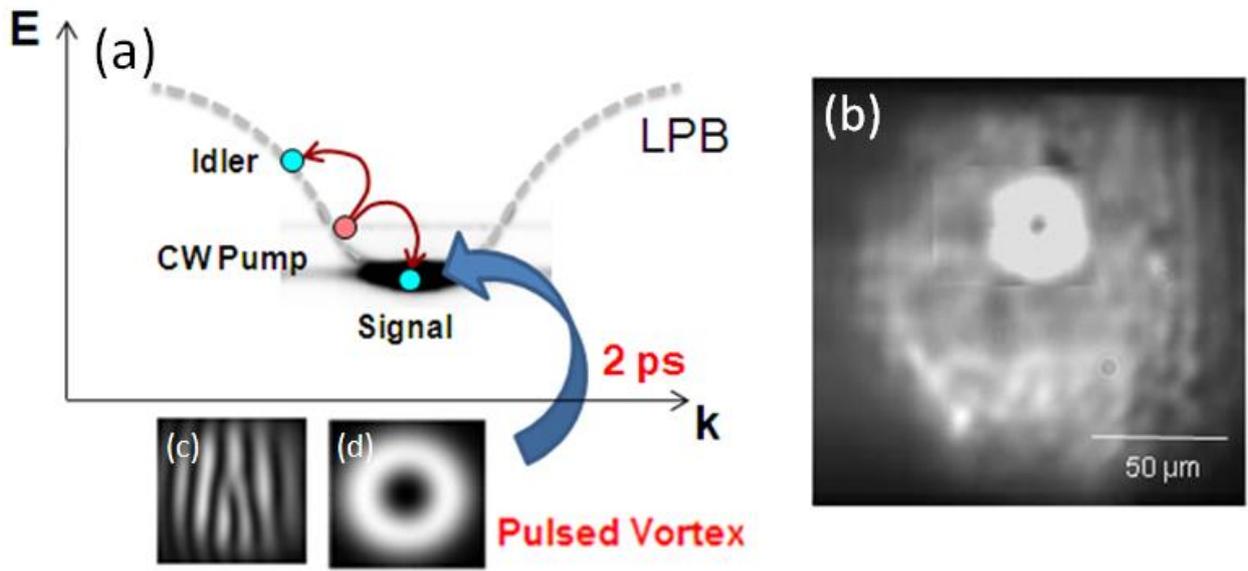

**FIGURE 2**

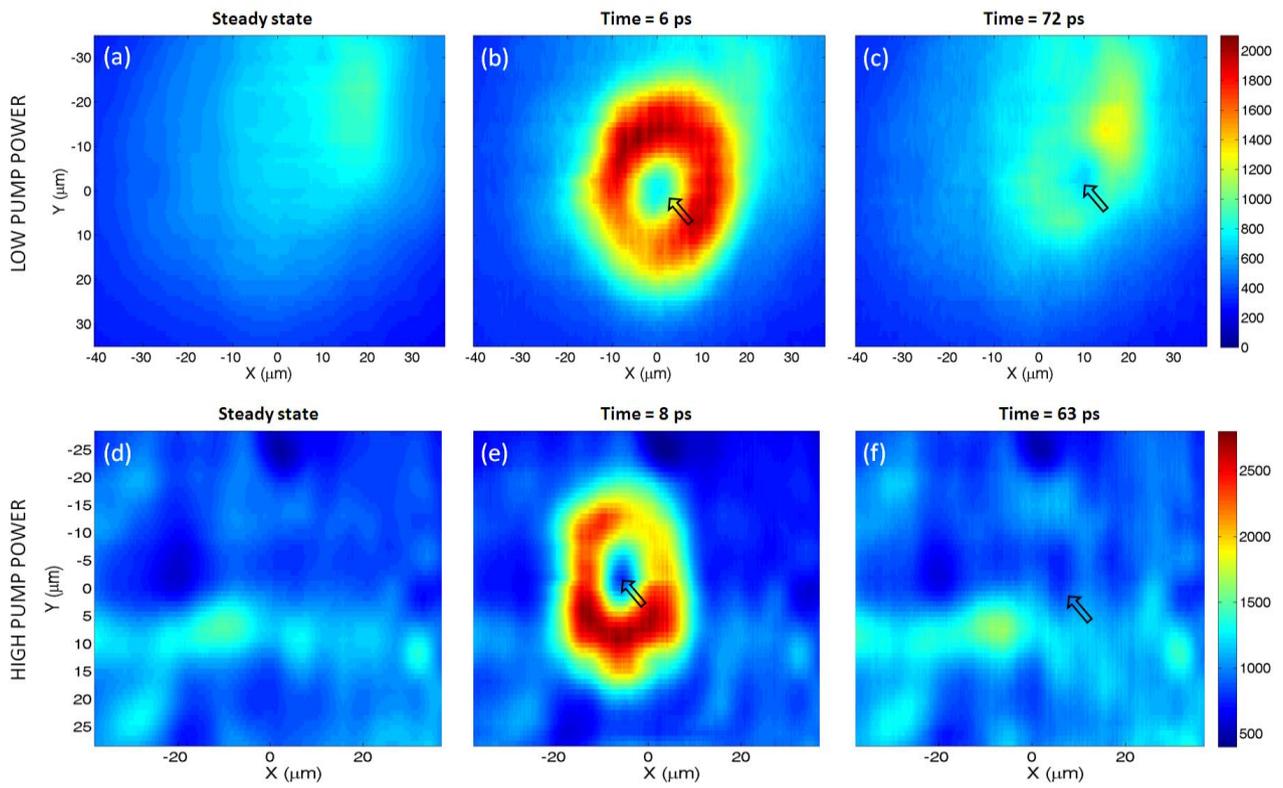

**FIGURE 3**

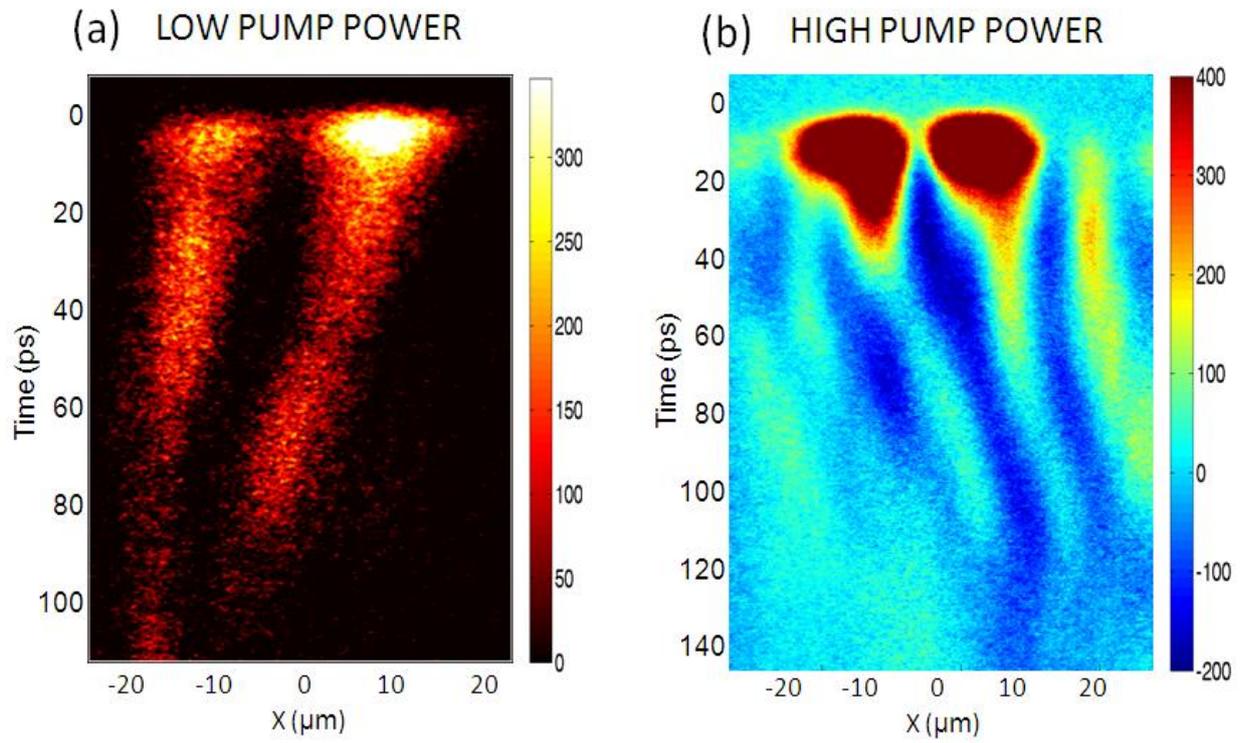

**FIGURE 4**

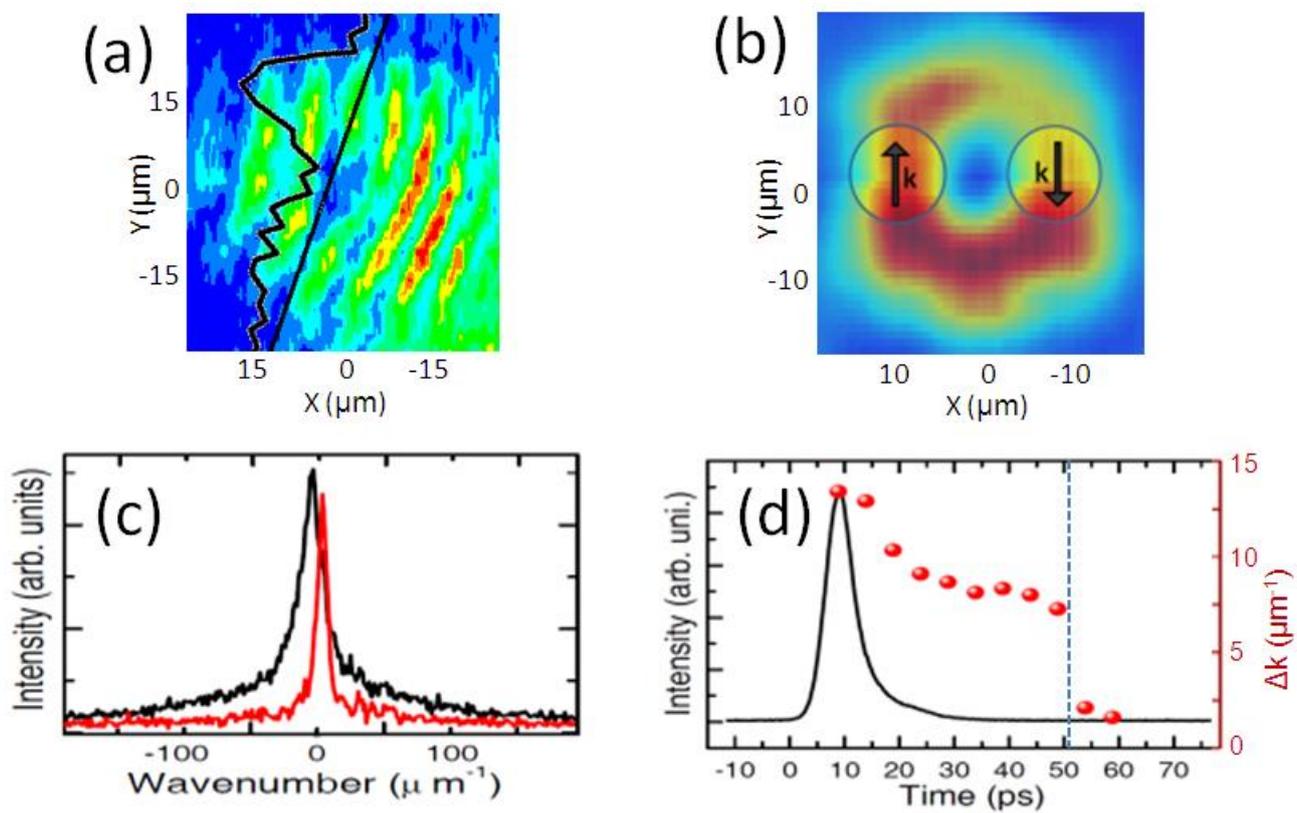

**FIGURE 5**

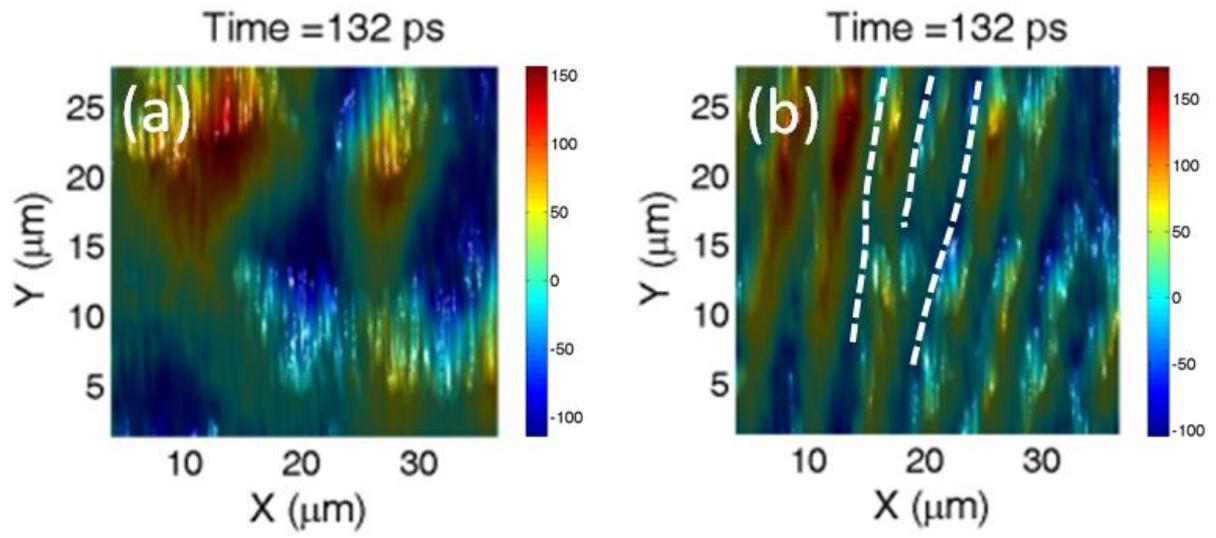

**FIGURE 6**

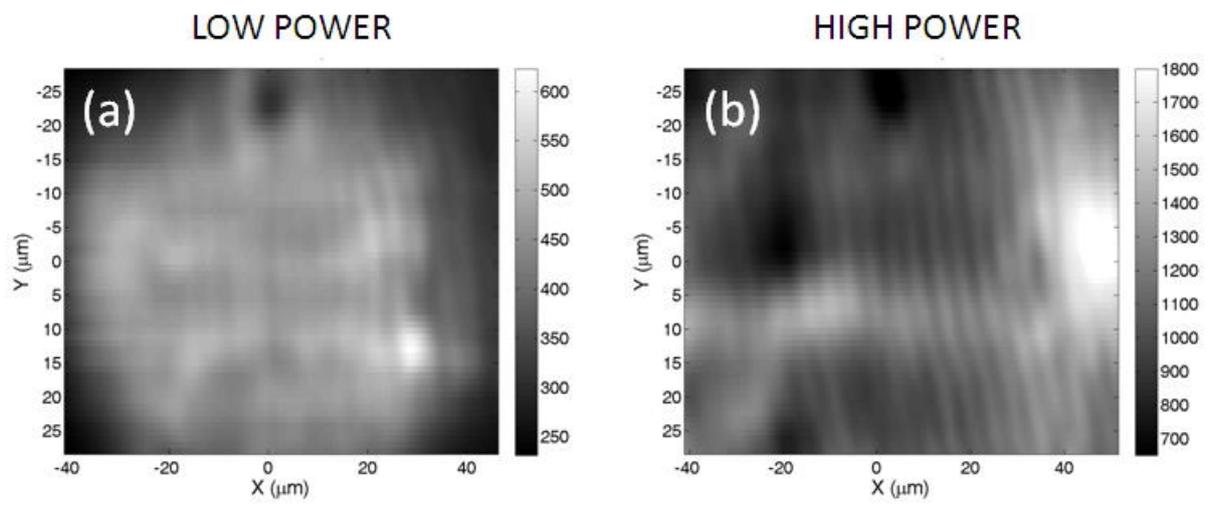